\documentclass[12pt,fleqn]{article}
\usepackage{graphics,epsfig,amsfonts, amssymb,amsmath}
\usepackage[english]{babel}
\newcommand{\be}{\begin{equation}}
\newcommand{\ee}{\end{equation}}
\newcommand{\beq}{\begin{eqnarray}}
\newcommand{\eeq}{\end{eqnarray}}
\newcommand{\beqn}{\begin{eqnarray*}}
\newcommand{\eeqn}{\end{eqnarray*}}
\newcommand{\ba}{\hspace*{-5pt}\begin{array}}
\newcommand{\ea}{\end{array}}
\newcommand{\p}{\partial}
\newcommand{\iy}{\infty}
\newcommand{\fr}{\frac}

\newcommand{\lb}{\label}

\newcommand{\ii}{\item}
\newcommand{\vsp}{\vspace{5mm}}
\newcommand{\bit}{\begin{itemize}}
\newcommand{\eit}{\end{itemize}}
\newcommand{\ben}{\begin{enumerate}}
\newcommand{\een}{\end{enumerate}}

\begin{document}

\begin{center}

{\bf On the stability of  kink-like and soliton-like solutions to
the generalized convection-reaction-diffusion equation}

\vsp

{\it V.A. Vladimirov and Cz. M\c{a}czka}

\vsp

Faculty of Applied Mathematics, \\
  AGH University of Science and Technology, \\
Mickiewicz Avenue 30, \\ 30-059 Krak\'{o}w, PL

\vsp

 E-mail: {\it vsevolod.vladimirov@gmail.com}

\end{center}

{\bf Abstract. }{
 Stability of the kink-like and soliton-like travelling wave solutions
to the generalized convection-reaction-diffusion equation is studied
by means of the qualitative methods and numerical simulation.}

\vsp

{\bf Keywords:} {\it generalized transport equation, active media,
temporal non-locality, traveling wave solutions, stability of wave
patterns}

\section{Introduction}

In this work we discuss the stability of traveling wave (TW)
solutions of the following family of convection-reaction-diffusion
equations: \be\lb{GBE}
\tau\,u_{tt}+u_t+g(u)\,u_{x}=\left[\kappa(u)\,u_x \right]_{x}+f(u).
\ee  Here $\tau\geq 0,$ $f(u)$ and $g(u)$  are peace-wise continuous
functions, $\kappa(u)$ is smooth and positive for
 $u>0$. Equations belonging to this family attracted attention of many
authors. The case $\tau=0$, recognized as
convection-reaction-diffusion (CRD) equation, is the subject of
investigations in several monographs
\cite{Barenblatt,Galaktionov,Maslov,Kersner}. It is widely used to
describe transport phenomena in porous media \cite{Richards}, theory
of combustion and detonation \cite{Zeldovich}, and mathematical
biology \cite{KPP,Murray}.

Besides the various applications, the CRD equation is valuable as
the simplest nonlinear model of transport phenomena having in some
cases nontrivial symmetry and, thus, possessing many exact solutions
and conservation laws
\cite{Kawahara,Galaktionov2,Clarkson-Mansfields,Cherniha,
Bar-Yur,Nik-Bar,Ivanova}.

General concepts leading to the family (\ref{GBE}) with $\tau\neq 0$
can be found in papers
\cite{Joseph-Preziosi,Makarenko,Makarenko-Moskalkov,Kar}. It can be
formally introduced if one changes in the balance equation the
convenient Fick's law
\[
J(t,\,x)=-\nabla\,q(t,\,x),
\]
stating the thermodynamical flow-force relations, with the
Cattaneo's equation
\[
\tau\fr{\p\,J(t,\,x)}{\p\,t}+ J(t,\,x)=-\nabla\,q(t,\,x),
\]
which takes into account the effects of memory.

Physically meaningful  TW solutions to Eq. (\ref{GBE}), such as
periodic, kink-like and, soliton-like solutions, compactons, shock
fronts, cuspons, and many other are either shown to exist or exactly
constructed in papers
\cite{VladKu_04,VladKu_05,VladKu_06,VlaMacz_07,Kar,Kutafina_10,Vladimir_10}.

The aim of this work is to analyze the stability of kink-like and
soliton-like TW solutions to Eq. (\ref{GBE}). The structure of the
work is following. In section 2 a geometric insight into the family
of TW solutions is made, and the linearized equations describing the
evolution of small perturbations of TW solutions, taking the form of
spectral problem are derived. In section 3 some important properties
of the continuous spectrum are stated, and the stability of the
constant asymptotic solutions is studied. In section 4 some
statements concerning the discrete spectrum are formulated. In
section 5 the results of numerical study of the temporal evolution
of the kink-like and soliton-like TW solutions, confirming and
supplementing the results of qualitative studies, are presented.

\section{Geometric insight into the TW solutions, and the statement of the problem}

This work is devoted to the analysis of stability the TW  solutions
having the form \be\lb{twsol} u(t,\,x)=U(z), \quad z=x-s\,t,\ee
where $s$ stands for the velocity of the traveling wave. To begin
with, let us make the geometric interpretation of some of the TW
solutions. Substituting the ansatz (\ref{twsol}) into the source
equation, we obtain the following ordinary differential equation:
\be\lb{SODE} \Delta U''(z)=\dot{\kappa}(U)\,{U'}^2+\left[
s-g(U)\right]\,{U'}+f(U), \ee where $\Delta=\tau\,s^2-{\kappa}(U)$,
both symbols $\dot{(\cdot)}$, and ${(\cdot)'}$  stand for the
derivative. This equation can be presented in the form of the
dynamical system \be\lb{Factor} \ba{l} \Delta\,
U'(z)=\Delta\, W(z), \\
\Delta\,W'(z)=\dot{\kappa}(U)\,W^2+\left[ s-g(U)\right]\,W+f(U). \ea
\ee

Now let us formulate a  simple statements concerning the dynamical
system (\ref{Factor}) and the properties of its solutions. \ben

\ii The stationary points of the system (\ref{Factor}) belong to
either the horizontal axis, or singular line $\Delta=0.$

\ii

\begin{figure}
\begin{center}
\includegraphics[width=5 in, height=1.5 in]{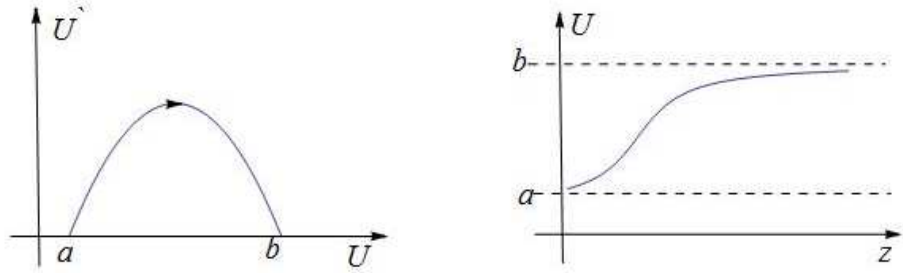}
\caption{Left: the heteroclinic trajectory in the phase plane
$(U,\,U')$; right: the corresponding kink-like solution
$U(z)$}\label{Fig:kink}
\end{center}
\end{figure}

A smooth kink-like solution is represented in the phase plane
$(U,\,\,W)$ by the heteroclinic trajectory, which does not intersect
the singular line $\Delta=0$ (see Fig. \ref{Fig:kink})

\ii

\begin{figure}
\begin{center}
\includegraphics[width=5 in, height=2.25 in]{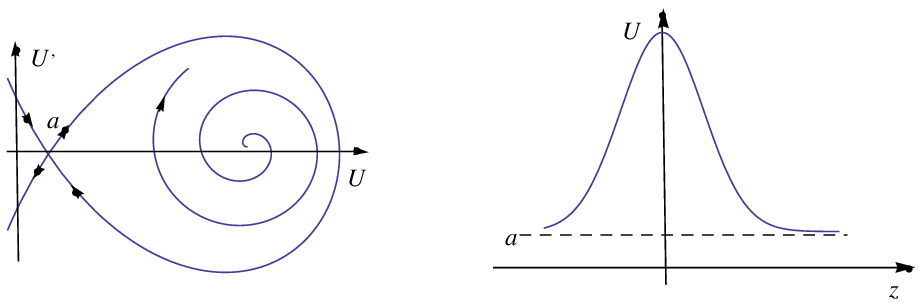}
\caption{Left: the homoclinic trajectory in the phase plane
$(U,\,U')$; right: the corresponding soiton-like solution
$U(z)$}\label{Fig:solit}
\end{center}
\end{figure}

The soliton-like solution is represented by the trajectory
bi-asymptotic to a saddle point which does not belong to the
singular line $\Delta=0$ (see Fig. \ref{Fig:solit}).

\ii

The trajectory bi-asymptotic to a saddle point lying in the singular
line $\Delta=0$ can correspond to either compacton, or shock front
(fore more detail see \cite{Vladimir_10,Vladimir_09a}).

\een

\begin{figure}
\begin{center}
\includegraphics[width=2.in, height=1. in]{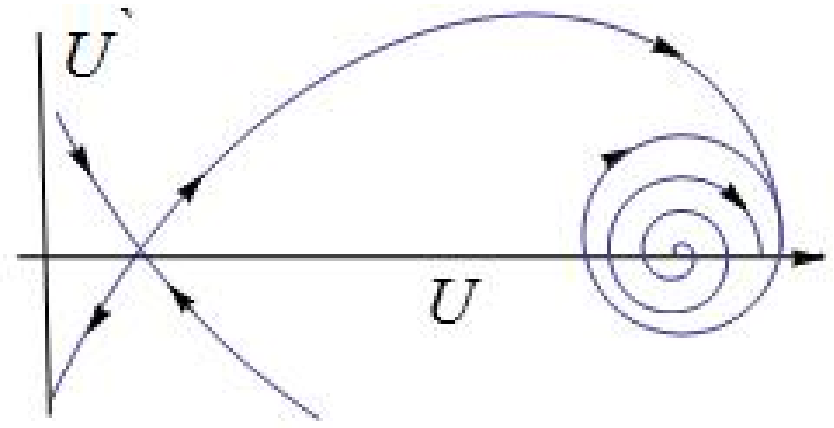}
\vspace{5mm}
\includegraphics[width=2.in, height=1. in]{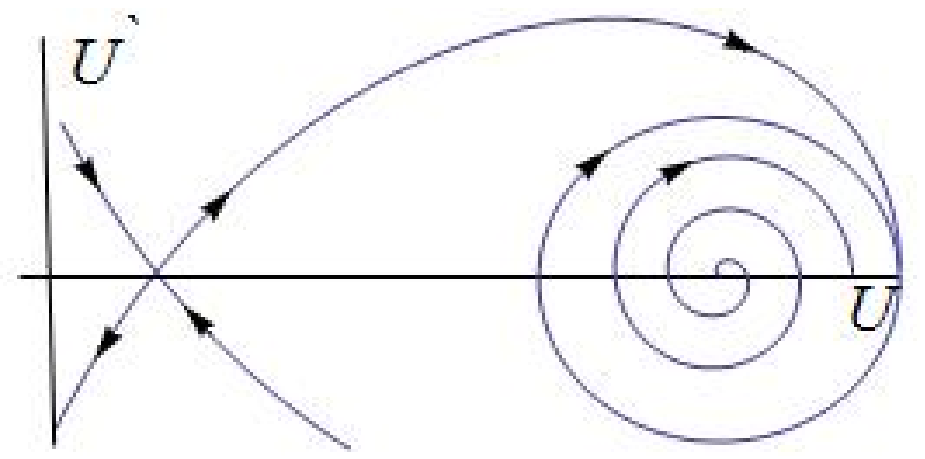}
\vspace{5mm}
\includegraphics[width=2.in, height=1. in]{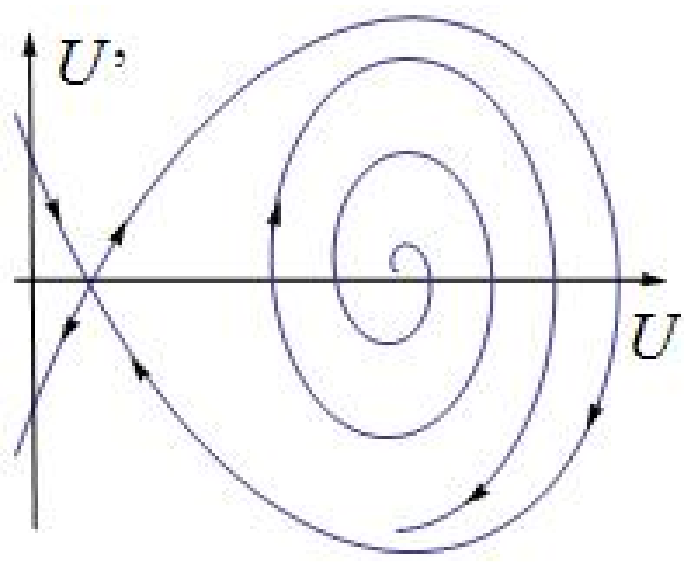}
\caption{Birth of the bi-asymptotic trajectory. Left: periodic
trajectory created as a result of the Hopf bifurcation; middle: the
change of driving parameter causes the grows of the radius of
periodic trajectory, which is finally destroyed  as a result of the
interaction with the nearby saddle point (right) }\label{Fig:HCLbif}
\end{center}
\end{figure}

Let us note that obtaining the analytical description to solitons,
compactons, or shock fronts in case of a typical dissipative system
like (\ref{Factor}) is rather difficult. But it is much more easy to
"capture" the homoclinic trajectory through two bifurcations: the
Hopf bifurcation, followed by the homoclinic bifurcation (see Fig.
\ref{Fig:HCLbif}). The Hopf bifurcation can be predicted by means of
the local asymptotic analysis \cite{GH, Has}. The homoclinic
bifurcation is nonlocal, and therefore should be  captured
numerically \cite{Vladimir_09,Vladimir_10}.

A simple analysis shows, that the equation (\ref{GBE}) can have the
homoclinic or heteroclinic TW solutions, if the corresponding
dynamical system possess at least two stationary points. This, in
turn, determines the form of the source term $f(u)$, which in the
simplest case is as follows: \be\lb{simple_f}
f(U)=(U-U_1)\,(U-U_0)\,\Psi(U). \ee We assume that $U_0\,<\,U_1$,
and $\Psi(U)$ does not intersect the horizontal axis within the
interval $(U_0,\,U_1).$

Let us now concentrate upon the problem of stability of the TW
solution (\ref{twsol}). Since we are interested in studying the
stability of kink-like and  soliton-like solutions, then we assume,
that \be\lb{asympt} \lim\limits_{z\to \pm \infty}\, U(z)=m_{\pm},
\ee where $m_{\pm}$ are the constants coinciding  with $U_0$ or
$U_1$ (in the case of a kink-like solution $m_{+}\,\neq\,m_{-}$, in
the case of soliton-like solution $m_{+}\,=\,m_{-}=U_0$). We assume
in addition, that \be\lb{lim_der} \lim\limits_{z\to \pm \infty}\,
U^{(k)}(z)=0 \ee for any natural $k$.

To study the stability of a TW solution $U(z),$ we use the ansatz
\be\lb{stab_ans} u(t,\,x)=U(z)+\epsilon\,\exp{[\lambda\,t]}\,v(z),
\ee where $\lambda$ is the spectral parameter, and  $|\epsilon|\ll\,
1.$ It is instructive to pass to new independent variables
\[
\bar{t}=t, \qquad \bar{z}=x-s\,t,
\]
in which the invariant solution (\ref{twsol}) becomes stationary. In
the new variables the equation (\ref{GBE}) reads as follows:
\begin{equation}\label{GBE1B}
\tau\,\left[\frac{\partial}{\partial\,\bar{t}}-s\frac{\partial}{\partial\,\bar{z}}\right]^2\,u+
\left[\frac{\partial}{\partial\,\bar{t}}-s\frac{\partial}{\partial\,\bar{z}}\right]\,u+
g(u)\,\frac{\partial\,u}{\partial\,\bar{z}}-\frac{\p}{\p\,\bar{z}}\left[\kappa(u)\,\frac{\partial\,u}{\partial\,\bar{z}}\right]=f(u)
\end{equation}
 (for simplicity, we omit the bars over the independent variables henceforth).
Up to $O(\epsilon^2)$, the function $v(z)$ satisfies the equation
\beq\lb{vareq} L\,\left[z,\,\fr{d}{d\,z},\,\lambda
\right]\,v(z)=\left\{\Delta\,\fr{d^2}{d\,z^2}+\left[g(U)-2\,\dot{U}\,\dot{\kappa}(U)-s(1+2\,\tau\,\lambda)
\right]\,\fr{d}{d\,z}+\right.  \\
\left.+\lambda(1+\tau\,\lambda)+\dot{g}(U)\dot{U}-\dot{\kappa}(U)\,\ddot{U}-\ddot{\kappa}(U)\dot{U}^2-\dot{f}(U)
\right\}\,v(z)=0,\nonumber \eeq where
$\Delta\,=\tau\,s^2-\kappa(U)$.

\vsp

{\bf Definition 1.} {\it The set of all possible values of
$\lambda\,\in\,\mathbb{C}$ for which the variational equation
(\ref{vareq}) has nontrivial solutions is called the spectrum of the
operator $L\,\left[z,\,\fr{d}{d\,z},\,\lambda \right]$.}

\vsp

{\bf Definition 2.} {\it We say that the TW solution $U(z)$ is
(linearly) stable, if any possible eigenvalue $\lambda$ for which
the equation (\ref{vareq}) has nonzero solution satisfies the
condition $\lambda\,\in\,\mathbb{C}\cup\,{0}$.}

\vsp

{\bf Remark.} It is easily seen, that zero eigenvalue always belongs
to the spectrum of the operator $L\,\left[z,\,\fr{d}{d\,z},\,\lambda
\right]$, for the following statement holds:

{\bf Lemma.} {\it If $U(z)$ is a TW solution of the equation
(\ref{GBE}), then $v_{\nu}(z)=\dot{U}(z)$ is the eigenvector of the
operator $L\,\left[z,\,\fr{d}{d\,z},\,0 \right]$.}

{\bf Proof.} Differentiating (\ref{SODE}) w.r.t. $z$, we can rewrite
the resulting equation in the form: \beqn
\left\{\left[\tau\,s^2-\kappa(U)\right]\,\fr{d^2}{d\,z^2}+\left[g(U)-2\,\dot{U}\,\dot{\kappa}(U)-s
\right]\,\fr{d}{d\,z}+\right. \\
\left.+\dot{g}(U)\dot{U}-\dot{\kappa}(U)\,\ddot{U}-\ddot{\kappa}(U)\dot{U}^2-\dot{f}(U)
\right\}\,v_\nu(z)=0, \eeqn where $v_\nu(z)$ stands for
$\dot{U}(z).$ The differential operator inside the braces coincides
with $L\,\left[z,\,\fr{d}{d\,z},\,\lambda \right]$, when
$\lambda=0.$ $\Box$

As usually, we distinguish the continuous spectrum
$\sigma_{cont}\,\subset\,\mathbb{C},$ and the discrete spectrum
$\sigma_{discr}\,\subset\,\mathbb{C}.$ Being somewhat informal, we
can treat $\sigma_{cont}$ as the subset responsible for the
stability of the stationary solutions $m_{\pm}$, and
$\sigma_{discr}$ for the stability of the solution $U(z)$ itself.

\section{Stability of the asymptotic stationary solutions}

Now we are going to state the conditions which guarantee that
$\sigma_{cont}\,\subset\,\mathbb{C}^{-}$. For this purpose, we study
the stability of the stationary solution $U_k=\mbox{const}$,
assuming that it coincides with  $m_{+}$ or $m_{-}$. We assume in
addition that the eigenvectors of the operators
$L\,\left[\pm\,\iy,\,\fr{d}{d\,z},\,\lambda \right]$ belong to the
space of tempered distributions $\mathcal{S'(R)}$ \cite{Maurin}.
With this assumption, we can solve the spectral problem
\be\lb{spinf} L\,\left[\ell,\,\fr{d}{d\,z},\,\lambda \right]v(z)=0,
\ee where $\ell$ stands for plus or minus infinity, by applying to
this equation the Fourier transformation. Thus, assuming that
$\lim\limits_{z\to \ell} U(z)=U_k,$ we get: \[ \ba{l}
F\left\{L\,\left[\ell,\,\fr{d}{d\,z},\,\lambda \right]\,v\right\}
[\xi]=
\\=L\,\left[\ell,\,-i\,\xi,\,\lambda \right]\,F[v](\xi)= \\
=\left\{-\left[\tau\,s^2-\kappa(U_k)
\right]\,\xi^2-i\,\xi\,\left[g\left(U_k\right)-s\,(1+2\,\lambda\,\tau)
\right]+\lambda\,(1+\lambda\,\tau)-\dot{f}\left(U_k\right)\right\}\,F[v](\xi)=0.\ea
\] This equation has nonzero solution $F[v](\xi)\,\in
\,\mathcal{S'(R)}$ if \be\lb{lam1} \lambda=\fr{-(1+2
i\,s\,\tau\,\xi)\pm\sqrt{Q}}{2\,\tau},
 \ee
where
\[
Q=1-4\tau\kappa\xi^2+4\tau\dot{f}(U_k)+4\,i\,\tau\xi\,g\left(U_k
\right).
\]
Since we are going to estimate the real part of $\lambda$, it is
instructive to use the representation
\[
\sqrt{Q}=x+i\,y, \quad \mbox{with}\quad x,\,\,y\,\in \mbox{R}.
\]
Raising both sides of this equation to the second power, and
eliminating $y$, we get the bi-quadratic equation
\[
x^4-x^2\left[1- 4 \tau\kappa\xi^2+4\tau\dot{f}(U_k)  \right]-4
\tau^2\xi^2g(U_k)^2=0.
\]
Its positive root takes the form
\[
x=\fr{1}{\sqrt{2}}\,\left[1-4\tau\kappa\xi^2+4\,\tau\dot{f}(U_k)+\sqrt{Q_1}
\right]^{1/2},
\]
where
\[
Q_1=\left(1- 4 \tau\kappa\xi^2+4\tau\dot{f}(U_k)\right)^2+16
\tau^2\xi^2g^2(U_k).
\]
So the biggest real part of $\lambda$, which we denote by
$Re\,\lambda^+$, is as follows:
\[
Re\,\lambda^+=\fr{1}{2\tau} \left\{-1+\fr{1}{\sqrt{2}}\left[
 1-4\tau\kappa\xi^2+4\,\tau\dot{f}(U_k)+\sqrt{Q_1}\right]^{1/2}\right\}.
\]
Let us solve the inequality $Re\,\lambda^+\,<\,0$ with respect to
$\xi\,\in\,{\mathbb{R}}$. It is equivalent to the inequality
\[
\left[
 1-4\tau\kappa\xi^2+4\,\tau\dot{f}(U_k)+\sqrt{Q_1}\right]^{1/2}\,<\,\sqrt{2},
\]
which, in turn, can be rewritten as
\[
\sqrt{Q_1}\,<\,1+4\tau\kappa\xi^2-4\,\tau\dot{f}(U_k).
\]
Rasing both sides of this inequality to the second power, we get,
after some algebraic manipulation, the inequality \be\lb{stabrelinf}
\xi^2\left[\tau g^2(U_k)-\kappa(U_k) \right]\,<\,-\dot{f}(U_k), \ee
which should be fulfilled for any $\xi\,\in \mathbb{R}.$ So the
following statement is true.

{\bf Statement1.} {\it The stationary solution  $U_k$ is stable if
\[
\tau g^2(U_k)-\kappa(U_k)\,<\,0, \quad \mbox{and} \quad
\dot{f}(U_k)<0.
\]
}

 In order to get $\sigma_{cont}$ in case $\tau=0$, we have to consider the  eigenvalue problem
\[\ba{l}
L_1\left[\ell,\,\fr{d}{d\,z}  \right]\,v(z)= \\
= \left\{ \kappa(U_k) \fr{d^2}{d\,z^2}+\left[s-g(U_k)
\right]\,\fr{d}{d\,z}+\dot{f}(U_k) \right\}\,v(z)=\lambda\,v(z).
 \ea\]
Applying the Fourier transformation, we get
\[
\lambda=i\,\xi\,\left[g(U_k)-s\right]-\kappa(U_k)\,\xi^2+\dot{f}(U_k),
\quad \xi\,\in\,\mbox{R}.
\]
So in the case $\tau=0$, we obtain the following result.

{\bf Statement 2.} {\it If  for $U_k=\lim\limits_{z\to\,\ell}U(z)$
$\dot{f}\left[U_k\right]<0$, $k=0,\,1$ then
$\sigma_{cont}\,\in\,\mathbb{C^-}$.}

\section{Some remarks concerning the discrete spectrum}

We remind that if the function $U(z)$ is the TW solution of the
equation (\ref{GBE}), then its derivative $\dot{U}(z)$ is the
eigenvector of the operator $L\left[z,\,\fr{d}{d\,z},\,0  \right]$,
corresponding to the eigenvalue $\lambda=0.$ This fact plays very
important rule when  $\tau=0$,  for it is possible to address the
question of stability of soliton-like and kink-like solutions by
employing the classical Sturm-Liouville theory \cite{Simon}. Let us
shortly remind one of its conclusions.

{\bf Theorem} (Sturm Oscillation Theorem). {\it Let
$\lambda_0>\lambda_2>...$ be the eigenvalues of the spectral problem
\[
\left\{H\left[z,\,\fr{d}{d\,z}\right]\right\}\,u(z)=\left\{\fr{d^2}{d\,z^2}+V(z)\right\}\,u(z)=\lambda\,u(z),
\qquad \lim\limits_{z\to A}u(z)=\lim\limits_{z\to B}u(z)=0,
\]
where  $u(z)\,\in\,L^2(A,\,B)$, $(A,\,B)\in\,\mathbb{R}$,  is finite
or infinite interval, $V(z)$ is a bounded function. Then the
eigenvector $u(z,\,\lambda_n)$ has exactly $n$ zeroes in $(A,\,B)$.}

Let us consider the eigenvalue problem for the case $\tau=0$,
assuming in addition that $\kappa=\mbox{const}.$ Then we can rewrite
the variational equation (\ref{vareq}) in the form \be\lb{var0}
\left[\fr{d^2}{d\,z^2}+a(z)\fr{d}{d\,z}+b(z)\right]\,v(z)=\tilde\lambda\,v(z),
\ee where $\tilde\lambda=\lambda/\kappa,$
\[
a(z)=\fr{s-g(U)}{\kappa}, \qquad
b(z)=\fr{\dot{f}(U)-\dot{U}\,\dot{g}(U)}{\kappa}.
\]
The problem (\ref{var0}) can be presented in the standard
Sturm-Liouville form, if we use the following transformation:
\be\lb{SL_form} v(z)=\exp\left[ \varphi(z) \right]\,w(z), \qquad
\mbox{where} \qquad \dot{\varphi}(z)=-\fr{a(z)}{2}. \ee Using
(\ref{SL_form}), we  obtain the eigenvalue problem \be\lb{SL}
\left[\fr{d^2}{d\,z^2}+\Phi(z)\right]\,w(z)=\tilde\lambda\,w(z), \ee
where
\[
\Phi(z)=b(z)-\fr{a^2(z)}{4}-\fr{a(z)}{2}.
\]

\begin{figure}
\begin{center}
\includegraphics[width=3.in, height=1.8 in]{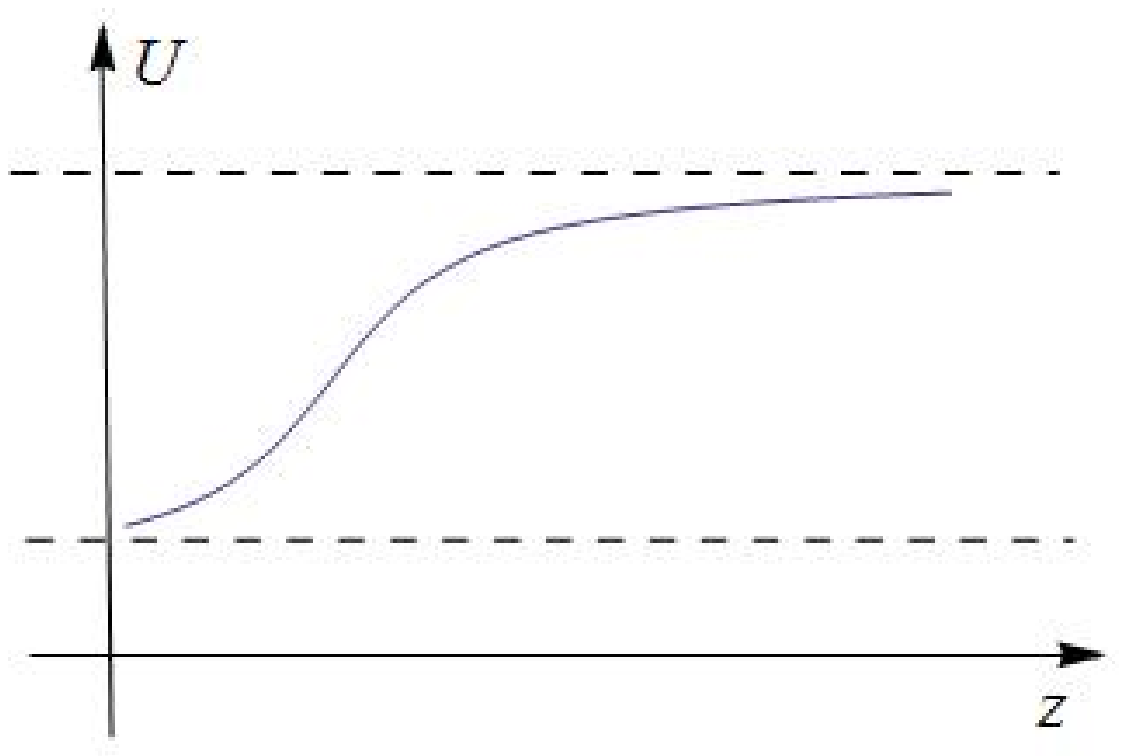}
\vspace{5mm}
\includegraphics[width=3.in, height=1.8 in]{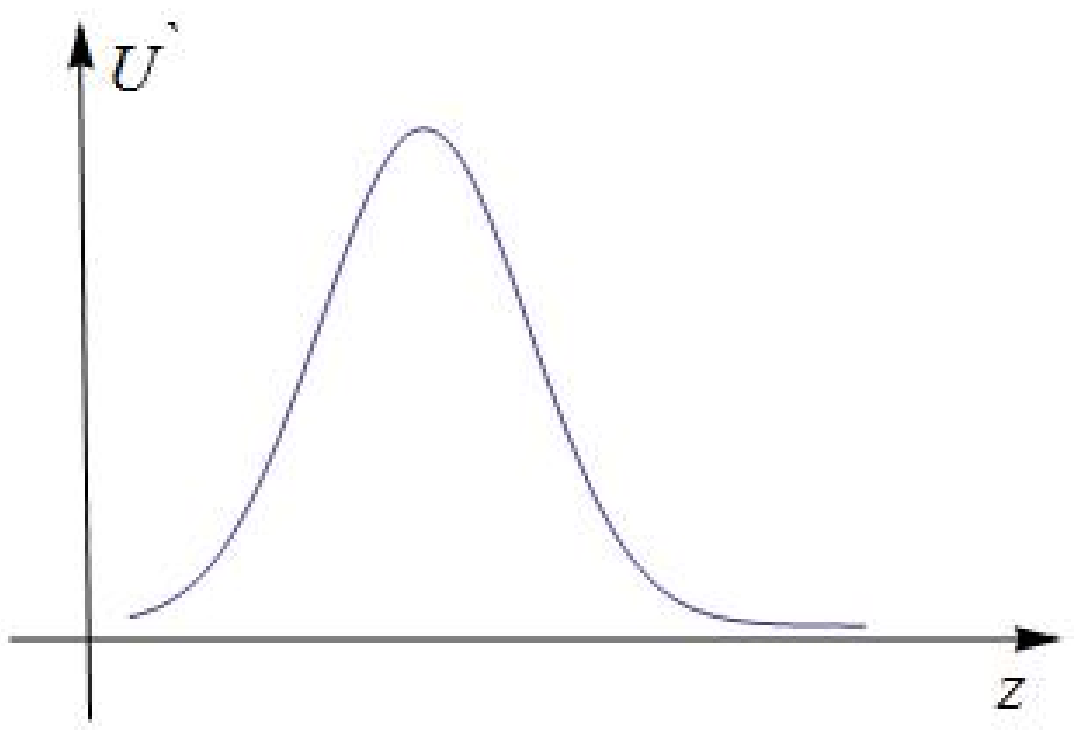}
\caption{Graphs of the monotonic kink-like solution (left), and its
derivative (right)}\label{Fig:kdot}
\end{center}
\end{figure}

\noindent By $w_{\nu}(z)$ we denote the eigenvector corresponding to
the eigenvalue $\tilde\lambda_\nu=0.$ And now, if $U(z)$ is the kink
with the monotone profile, then $v_\nu(z)=\dot{U}(z)$, as well as
 $w_\nu(z)=\dot{U}(z)\,\exp\left[-\varphi(z)\right]$,
is a functions, which does not intersect the real axis, see Fig
\ref{Fig:kdot}. Then on virtue of the Sturm Oscillation Theorem,
$\tilde\lambda_\nu=\tilde\lambda_0=0$, and all the remaining
eigenvalues are negative.

\begin{figure}
\begin{center}
\includegraphics[width=2.8 in, height=1.8 in]{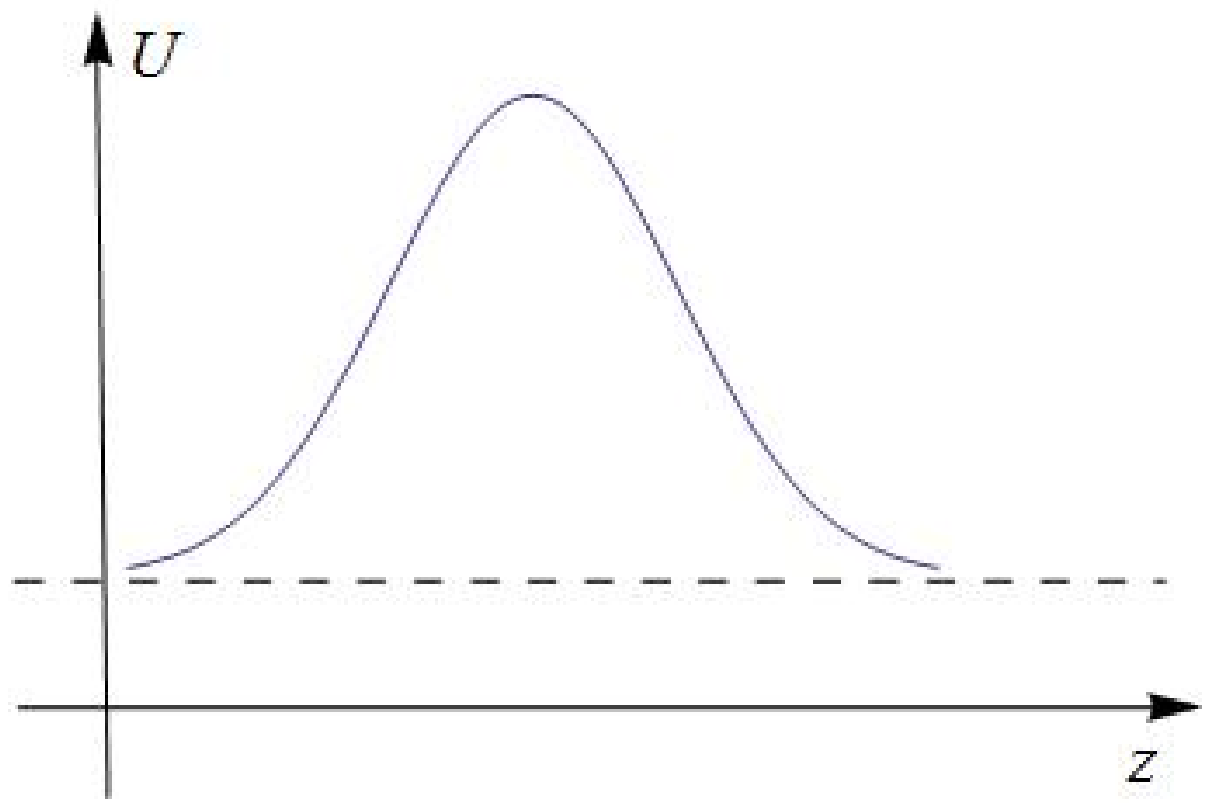}
\vspace{5mm}
\includegraphics[width=2.8 in, height=1.8 in]{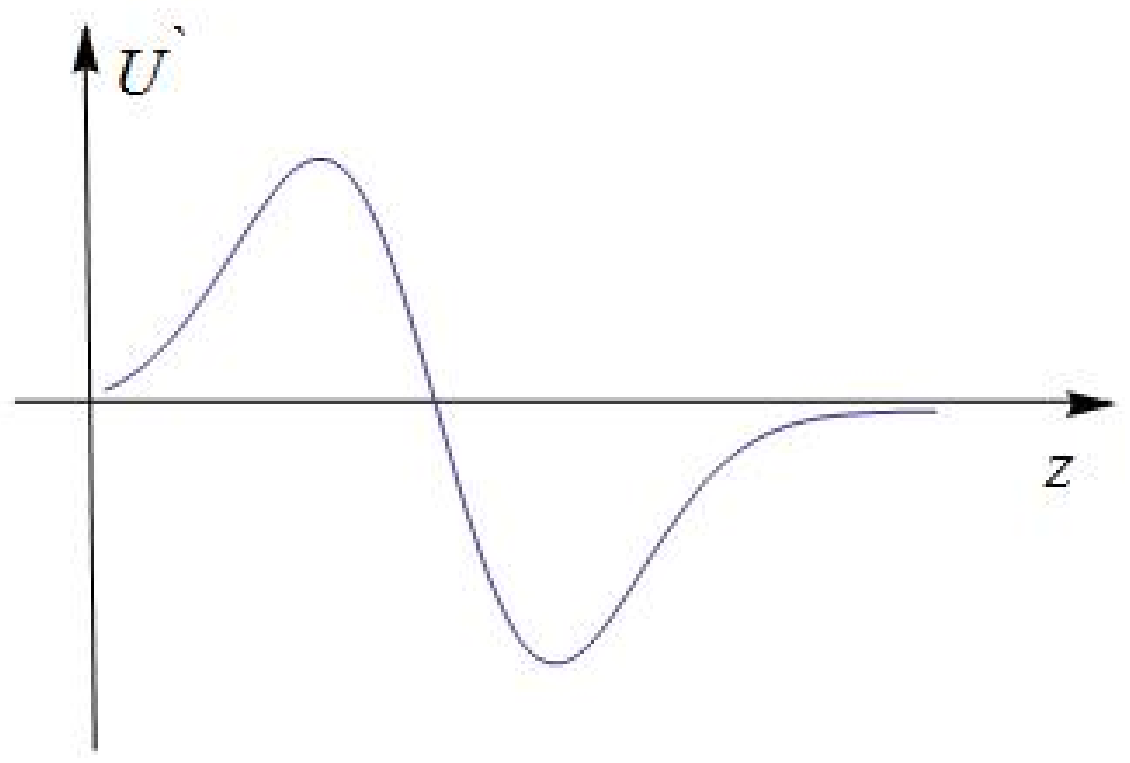}
\caption{The graphs of one-humped soliton-like solution $U(z)$
(left), and  its derivative $\dot{U}(z)$ (right)}\label{Fig:sdot}
\end{center}
\end{figure}

On the other hand, if $U(z)$ is the soliton-like TW solution, then
both  $v_\nu(z)=\dot{U}(z)$, and
$w_\nu(z)=\dot{U}(z)\,\exp\left[-\varphi(z)\right]$ intersect the
horizontal axis, see Fig \ref{Fig:sdot}. So
$\tilde\lambda_\nu=\tilde\lambda_1=0$, and there is the eigenvalue
$\tilde\lambda_0\,>0$  belonging to the right half-plane of the
complex plane. The latter result is rather well-known \cite{Idris}.
Conditions concerning the stability of kink-like solution can be
formulated as follows.

{\bf Statement 3.} {\it A monotonic kink-like solution of the
equation (\ref{GBE}) with $\tau=0$ is stable, provided that
$\dot{f}(U_k)<0$ for $k=0,\,1.$ }

It is possible to apply the  approach based on the Sturm-Liouvliie
theory  in the case when $\tau>0$, and  $g(u)$, $\kappa(u)$ are
constant functions. Under these conditions, the variational equation
can be presented in the form \be\lb{var1}
\left[\fr{d^2}{d\,z^2}+\tilde{a}(z)\fr{d}{d\,z}+\tilde{b}(z)\right]\,v(z)=\theta\,v(z),
\ee where
\[
\tilde{a}(z)=\fr{s\,(1+2\,\tau\,\lambda)-g}{\mu}, \quad
\tilde{b}(z)=\fr{\dot{f}(U)}{\mu}, \quad \theta=\fr{\lambda
(1+\tau\,\lambda)}{\mu},
\]
$\mu=\kappa-\tau\,s^2$, The transformation
\[
v(z)=\exp\left[ \varphi(z)\,w(z) \right], \qquad \mbox{where} \qquad
\dot{\varphi}(z)=\fr{g-s(1+2\,\tau\,\lambda)}{2\,\mu}
\]
leads in this case to the  spectral problem
\[
\left[\fr{d^2}{d\,z^2}+\Psi(z)\right]\,w(z)=\chi(\lambda)\,w(z),
\quad \lim\limits_{z\to a}w(z)=\lim\limits_{z\to b}w(z)=0,
\]
where
\[
\Psi(z)=\fr{g(2\,s-g)-s^2}{4\,\mu^2}+\fr{\dot{f}[U(z)]}{\mu}, \qquad
\chi(\lambda)=\fr{\lambda[\kappa-g\,s\,\tau+\kappa\,\tau\,\lambda]}{\mu^2}.
\]

\begin{figure}
\begin{center}
\includegraphics[width=2.5 in, height=1.5 in]{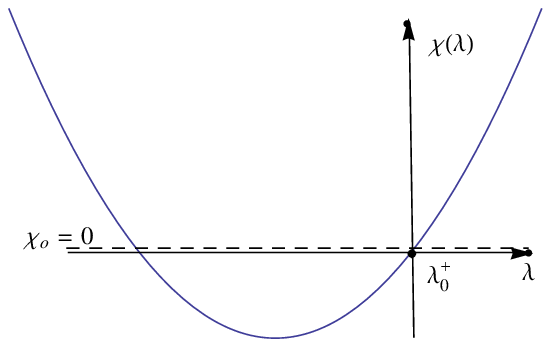}
\vspace{5mm}
\includegraphics[width=2.5 in, height=1.5 in]{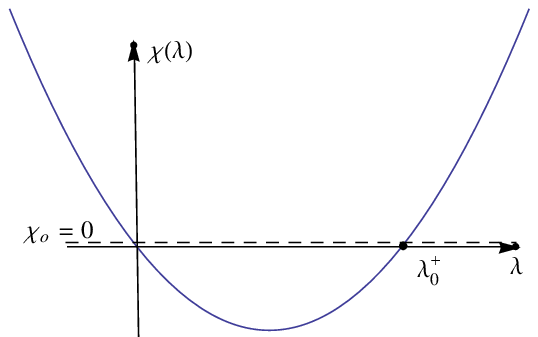}
\caption{Graphs of the function $\chi(\lambda)$ and the eigenvalues
$\lambda_0^+$ corresponding to $\dot{U}(z)$ when ${U}(z)$ is a
kink-like TW solution. Left: $\kappa-g\,s\,\tau>0$; right:
$\kappa-g\,s\,\tau<0$}\label{Fig:kinkroots}
\end{center}
\end{figure}

Thus, if $U(z)$ is a monotonic kink-like solution, then
$v_0(z)=\dot{U}(z)$ is the eigenvector corresponding to the
eigenvalue $\lambda=0.$ The corresponding function
$w_0(z)=e^{-\varphi(z)}\,v_0(z)$ is the eigenvector of the operator
$\hat{H}=\fr{d^2}{d\,z^2}+\Psi(z)$. Since the function $w_0(z)$ does
not intersect the horizontal axis, then, on virtue of the Sturm
Oscillation Theorem, it corresponds to the eigenvalue $\chi_0=0$,
and any other eigenvalue of this problem is negative. But
$\chi(\lambda)$ is the quadratic function of $\lambda$, and
therefore the source eigenvalue problem (\ref{var1}) can have an
extra eigenvalue, corresponding to the function $\dot{U}(z).$ This
extra root will be negative, if the constant $\kappa-g\,s\,\tau,$ is
positive, and negative otherwise, see Fig. \ref{Fig:kinkroots}.
Hence for $\tau>0$ we get an extra condition \be\lb{stabkink}
\kappa-g\,s\,\tau \,>\,0,\ee assuring that
$\sigma_{discr}\,\in\,C^-.$

\begin{figure}
\begin{center}
\includegraphics[width=2.5 in, height=1.5 in]{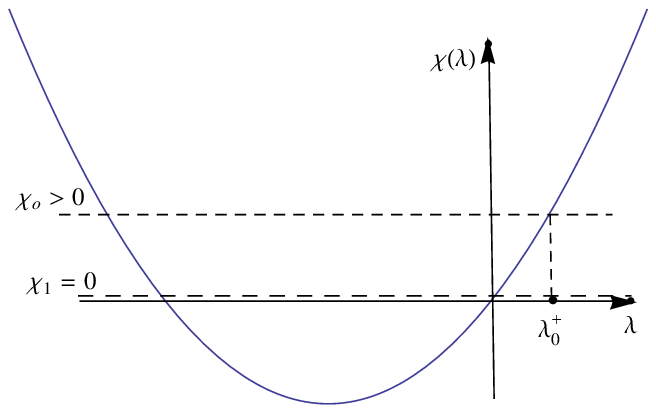}
\vspace{5mm}
\includegraphics[width=2.5 in, height=1.5 in]{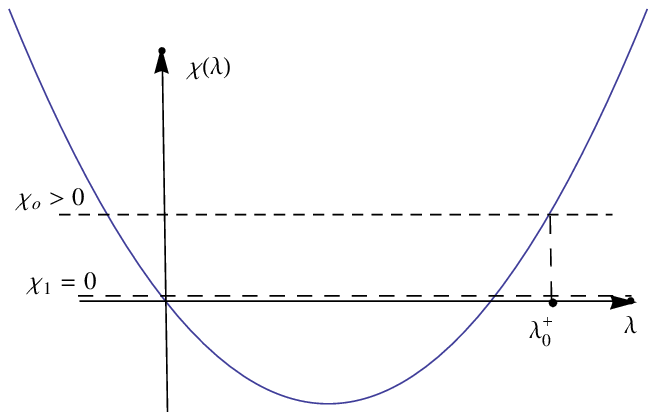}
\caption{Graphs of the function $\chi(\lambda)$ and the eigenvalues
$\lambda_0^+$ corresponding to $\dot{U}(z)$ when ${U}(z)$ is a
soliton-like TW solution. Left: $\kappa-g\,s\,\tau>0$; right:
$\kappa-g\,s\,\tau<0$}\label{Fig:solitroots}
\end{center}
\end{figure}

If, in turn, $U(z)$ is the soliton-like solution, then  the
eigenvector $w_{\nu}=\exp{[-\varphi(z)]}\,\dot{U}(z)$ corresponding
to the eigenvalue $\chi_\nu=0$ intersects once the horizontal axis.
Hence $\nu=1,$ and there is an extra eigenvalue $\chi_0>\chi_1=0$,
to which corresponds a pair of the eigenvalues $\lambda_0^{\pm}$ of
the source eigenvalue problem. As it is seen in Fig.
\ref{Fig:solitroots}, regardless of the sign of $\kappa-g\,s\,\tau$,
there exists the positive eigenvalue $\lambda_0^{+}$, hence the
soliton-like solution is unstable.

\section{The results of numerical simulations}

In the general case, i.e., when, e.g., the function $g(u)$ is not
constant, estimation of $\sigma_{discr}$ is rather  more delicate
problem. In paper \cite{VlaMacz_11} such estimation is performed for
the TW solution \be\lb{U123} u(t,\,x)=U(z)=\fr{\Psi'(z)}{\Psi(z)},
\qquad
\Psi(z)=\exp{[m_1\,z]}+C_2\,\exp{[m_2\,z]}+C_3\,\exp{[m_3\,z]}, \ee
satisfying the equation \be\lb{GBE1}
\tau\,u_{tt}+u_t+\mu\,u\,u_{x}=\kappa\,u_{x\,x}+\nu \,(u-m_1)\,
(u-m_2) (u-m_3), \ee under the following restrictions on the
parameters:
\[
\mu=3\,\Delta, \quad \nu=-\Delta, \quad
s=(m_1+m_2+m_3)\,\Delta\equiv\,\fr{1+\sqrt{1+4\,\tau\,\kappa
(\sum_{k=1}^3{m_k})^2}}{2\,\tau\,\sum_{k=1}^3{m_k}}
\]
Unfortunately, $\sigma_{cont}$  for none of these solutions is
contained in $\mathbb{C^-}$, as will  be shown below.

{\bf Statement 4.} {\it If $C_2^2+C_3^2>0$, and
$0\,\leq\,m_1<m_2<m_3$, then the stationary point
$m_{+}=\lim\limits_{z\,\to\,+\infty}U(z)$ is unstable.
  }

  {\it Proof.} The stationary point $U_k=m_{+}$ is stable, if the
  inequality (\ref{stabrelinf}) is fulfilled for all $\xi\in\,\mathbb{R}$. The
  case $C_3=0$ is very easy to analyze. Indeed, under this
  assumption $m_{+}$ coincides with $m_2$, and thus
  \[\dot{f}(m_2)=\nu\,(m_2-m_1)\,(m_2-m_3)=-\Delta\,(m_2-m_1)\,(m_2-m_3)>0.
  \]
Hence the conditions of the statement 1 are not fulfilled.

If $C_3\,\neq\,0$, then $\lim\limits_{z\,\to\,+\iy}=m_3$, and
\[
\dot{f}(m_3)=\nu\,(m_3-m_1)\,(m_3-m_2)=-\Delta\,(m_3-m_1)\,(m_3-m_2)<0.
\]
Let us address the condition
\[
\tau \,g^2(m_3)-\kappa\,=\,\tau \,(\mu\,m_3)^2-\kappa\,=\,\tau (3
\,\Delta\,m_3)-\kappa\,<\,0.
\]
This inequality is equivalent to the following one:
\[
\Delta\,=\,\fr{1+\sqrt{1+4\,\tau\,\kappa\,M^2}}{2\,\tau\,M^2}<\,\sqrt{\fr{\kappa}{9\,\tau\,m_3^2}},
\quad M=m_1+m_2+m_3,
\]
which can be rewritten as
\[
\sqrt{1+4\,\tau\,\kappa\,M^2}\,<\,2\,\tau\,M^2\sqrt{\fr{\kappa}{9\,\tau\,m_3^2}}-1.
\]
It is evident, that the above inequality cannot be fulfilled if the
RHS is negative, so let us assume, that
$2\,\tau\,M^2\sqrt{\fr{\kappa}{9\,\tau\,m_3^2}}-1>0.$ Raising both
sides to the second power, we get, after some algebraic
manipulation, the inequality \be\lb{ineq}
\sqrt{\fr{\kappa}{9\,\tau\,m_3^2}}<\fr{\kappa}{9\,m_3^2}\,\left(M^2-9\,m_3^2
\right). \ee The inequality (\ref{ineq}) cannot be fulfilled since
\[
M^2-9\,m_3^2=(m_1+m_2+m_3)^2-9\,m_3^2<\left(3\,m_3
\right)^2=9\,m_3^2=0.
\] $\Box$

\begin{figure}
 \centering\includegraphics[width=3 in, height=2 in ]{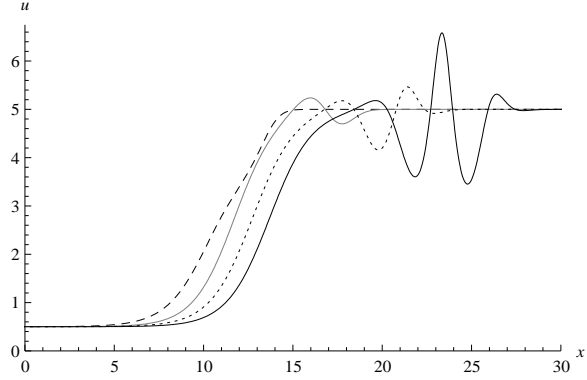}
\caption{Numerical solution  of the system (\ref{GBE1}) in case when
$\tau=\kappa=1,\, m_1=0.5,\, m_2= 1.5,\, m_3 = 5,\, C_2=1,\, C_3=3,
$ and the  invariant  kink-like solution (\ref{U123}) is taken as
the Cauch\'{y} data. Successive graphs present the TW, moving from
left to right}\label{Fig:Kinkdestruct}
\end{figure}

The above statement tells us, that, whether or not $\sigma_{discr}$
belongs to the left half-plane,  the traveling wave (\ref{U123})
cannot evolve in a  self-similar mode. Example presented in Fig.
\ref{Fig:Kinkdestruct} confirms this conclusion. The numerical
solution of the Cauchy problem with the Cauchy data
$u(0,\,x),\,\,u_t(0,\,x)$ being equal, respectively, to
$\Psi'(x)/\Psi(x)$ and $-s\,\left(\Psi'(x)/\Psi(x)\right)'$,
performed under the following values of the parameters $\tau=1
,\,\,\kappa=1, m_1=0.5,\,m_2=1.5,\,m_3=5,C_2=1,\,C_3=3$ shows that
the initial perturbation evolves for some time in a self-similar
mode. In the long run the self-similar evolution becomes corrupted.
The process of self-similarity destruction starts from the far end,
and this is in agreement with the observation that the condition
(\ref{stabrelinf}) is not fulfilled for $u=m_3=\mbox{const}$.

\begin{figure}
 \centering\includegraphics[width=3 in, height=2 in ]{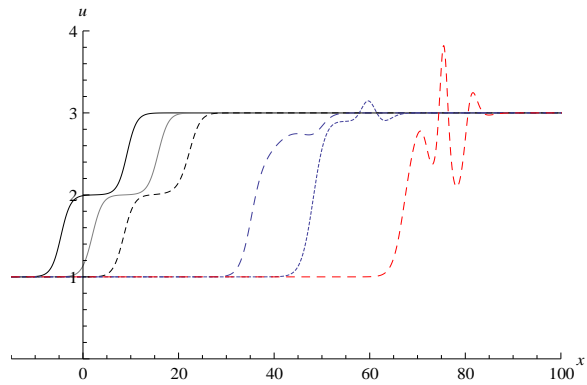}
\caption{Numerical solution  of the system (\ref{GBE1}) in case when
$\tau=\kappa=1,\,\, m_1=1,\, m_2= 2,\, m_3 = 3,\, C_2=100,\,
C_3=0.01, $ and the invariant  kink-like solution (\ref{U123}) is
taken as the Cauch\'{y} data. Successive graphs present the TW,
moving from left to right}\label{Fig:U123step}
\end{figure}

Numerical experiments performed with another values of the
parameters, namely: $\tau=1 ,\,\kappa=1,\,
m_1=1,\,m_2=2,\,m_3=3,C_2=100,\,C_3=0.01$ expose the same tendency,
Fig \ref{Fig:U123step}.

Numerical simulations purposed at studying the evolution of
soliton-like solutions to the equation (\ref{GBE}) show, that these
solutions are rather  unstable. In a number of numerical
simulations, performed with different functions $g(u)$ and
$\kappa(u)$, we had encountered three types of instabilities
destroying the TW solution. The first type is connected with the
instability of the constant asymptotic solution. Two other types are
manifested in either fading or blowing-up of solution in finite
time. Since this behavior is rather typical, let us illustrate it on
the example of the solitary wave solutions of the equation
\be\lb{duffinglike} \tau\,u_{tt}+u_t+u\,u_{x}=\kappa\,u_{x\,x}
+\gamma\, u\,\left(1-u^2\right) \ee considered in paper
\cite{VladKu_04}. The factorized system \be\lb{Factorduf} \ba{l}
\Delta\,
U'(z)=\Delta\, W(z), \\
\Delta\,W'(z)=\left[ s-U[z]\right]\,W+U[z]\,\left(1-U^2[z]\right),
\ea \ee obtained via the substitution (\ref{twsol}), possesses the
homoclinic solutions attained through two bifurcations. The first
one is the Hopf bifurcation taking place when $\gamma\,\Delta>0$
\cite{VladKu_05}, and the parameter $s$ is close to the unity. The
second one is non-local and is captured numerically.

In both possible cases, i.e., when for  $\gamma$ and $\Delta$ are
simultaneously positive or negative,  variation of velocity $s$ near
the unit value leads to the appearance of homoclinic loop
corresponding to the soliton-like solution.

\begin{figure}
\begin{center}
\includegraphics[width=3.in, height=2.25 in]{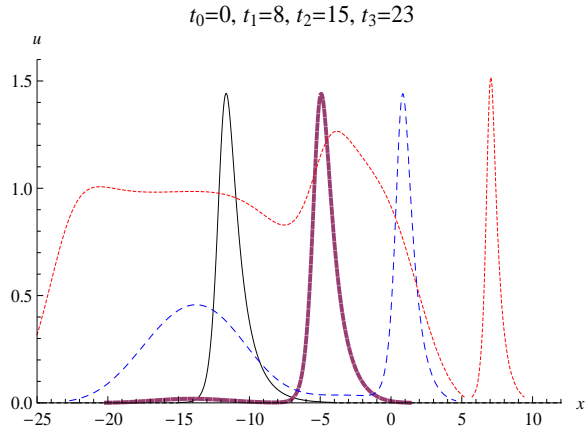}
\caption{Numerical solution  of the system (\ref{duffinglike})
performed with  $\tau=2,\,\kappa=1,\,\,\gamma=1$, when the invariant
soliton-like solution obtained by solving
 the system (\ref{Factorduf}) is taken as the
Cauch\'{y} data. }\label{Fig:Duffinglike1}
\end{center}
\end{figure}

\begin{figure}
\begin{center}
\includegraphics[width=3.in, height=2.25 in]{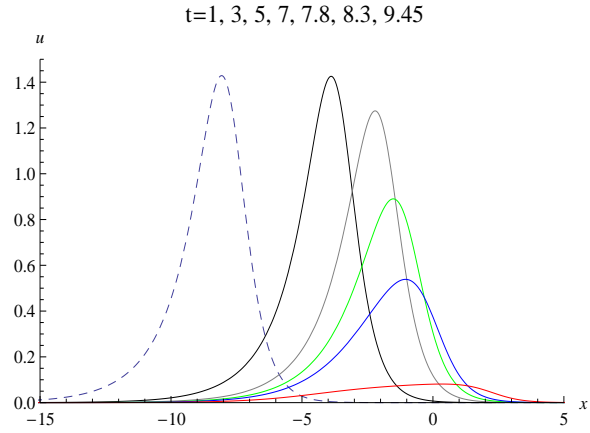}
\caption{Numerical solution  of the system (\ref{duffinglike})  with
$\tau=0.25,\,\kappa=1,\,\,\gamma=-1\,\,(M=0.71)$ when the invariant
soliton-like solution  obtained by solving
 the system (\ref{Factorduf}) is taken as the
Cauch\'{y} data.}
 \label{Fig:Duffingfading}
\end{center}
\end{figure}

\begin{figure}
\begin{center}
\includegraphics[width=3.in, height=2.25 in]{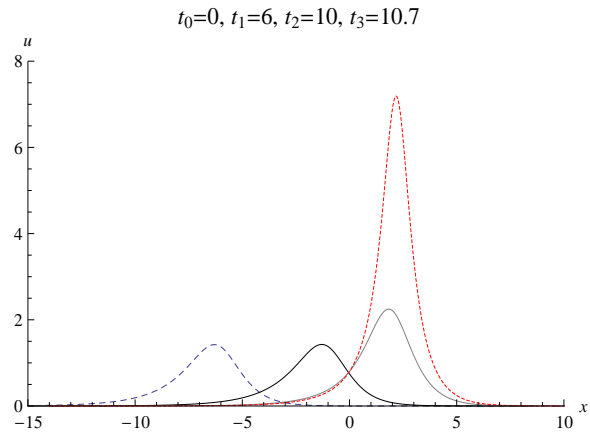}
\caption{Numerical solution  of the system (\ref{duffinglike}) in
case when the invariant  soliton-like solution  obtained by solving
 the system (\ref{Factorduf}) with $\tau=1,\,\kappa=2,\,\,\gamma=-1\,\,(M=1.004)$ is taken as the Cauch\'{y} data.}
 \label{Fig:Duffinglike2}
\end{center}
\end{figure}

Numerical integration of the Cauchy problem for (\ref{duffinglike})
with the soliton-like solution taken as the Cauchy data, performed
with the following values of the parameters $\tau=2,\, \kappa=1,\,
\gamma=1,\,\,s=0.84912$ (when both $\Delta$ and $\gamma$  are
positive), reveals that the initial solitary wave evolves in a
self-similar mode for some time,  but finally  is destroyed as a
result of instability of the asymptotic state, Fig.
\ref{Fig:Duffinglike1}. Let us note, that the instability of the
stationary asymptotic solution $u=U_0=0$ is confirmed by inspection
of the formula (\ref{stabrelinf}). Evolution of the soliton-like
solution appearing in two other cases depends on the magnitude of
the characteristic number $M={u_{max}}/{C_0}$, where $u_{max}$ is
the maximal amplitude of the initial perturbation,
$C_0=\sqrt{\kappa/\tau}$ is the acoustic waves' characteristic
velocity. We call $M$ the "Mach number", since it plays the
analogous role, as the parameter known under this name plays in the
theory of supersonic flows. Indeed, if $M<1$, then the initial
solitary wave vanishes to zero, Fig. \ref{Fig:Duffingfading}. Let us
note, that we deal in this case with the destruction mechanism,
which is completely different from the convenient dispersion. The TW
evolves for a while in self-similar mode, but at some instant its
amplitude is subjected to the drastic decrease so that the wave pack
completely vanishes in finite time.

When $M>1$ then the evolution of the solitary wave ends with the
blow-up regime appearance, Fig. \ref{Fig:Duffinglike2}. We would
like to mention in this place, that such a strong dependence  of
solutions of the nonlinear hyperbolic-type equation similar to
(\ref{GBE}) upon the "Mach number" was noted for the first time in
paper \cite{Makarenko-Moskalkov}.

\section{Discussion}

Thus in this work qualitative and numerical investigations of the TW
solutions of the system (\ref{GBE}) have been performed, with
special attention paid to the behavior of the kink-like and
soliton-like solutions. On the basis of the results obtained, we can
state that incorporation of  the second-derivative with respect to
time does not  lead to drastic change of  the situation taking place
in the case of the classical convection-reaction-diffusion equation,
for which a monotonic kink-like solution is stable if both of  the
constant asymptotic solutions are stable, while the soliton-like
solution is always unstable. In fact, the situation in the case of
the equation (\ref{GBE}) is not so clear, for the Sturm Oscillation
Theorem cannot be directly applied when $g(u)$ and $\kappa(u)$ are
nontrivial functions. Yet in the situation when both $\kappa(u)$ and
$g(u)$ are constant, and we can formulate the variational problem in
the Sturm-Liouville form, the results of qualitative analysis show,
that an extra inequality should be fulfilled in order that the
kink-like solution be stable. Under the same conditions, any
soliton-like solution proves to be unstable, as this is the case
when $\tau=0.$ Numerical experiments performed with different
$\kappa(u)$, $g(u)$ and $f(u)$ for which the source equation
possesses the  solitary-wave solutions (taken as the Cauchy data),
reveal the instability  in the wide range of the parameters' values.
They evidence that solitons, compactons, and shock waves appearing
in the class of the generalized convection-reaction-diffusion
equations remain unstable in the case of positive $\tau$.


\begin{thebibliography}{99}
\bibitem{Barenblatt}
Barenblatt G. I., {\em Similarity, Self-similarity and Intermediate
Asymptotics}, Consultants Bureau, New York, 1979.
\bibitem{Galaktionov}
Samarskii A., Galaktionov V.,Kurdiumov A., Mikhailov A., {\em
Blow-up in Quasilinear Parabolic Equations,} Walter de Gruyter, NY,
1995.
\bibitem{Maslov}  Danilov V., Maslov V., and Volosov K.,
{\em Mathematical Modelling of heat and Mass Transfre Processes,}
Kluver Academic Publ., Dordrecht, Boston, 1995.
\bibitem{Kersner}  Gilding B. H., Kersner R., {\em Travelling Waves in Nonlinear
Diffusion-Convection-Reaction}, Birkhauser, 2004.
\bibitem{Richards} Richards L. A., {\em Capillarity conduction of liquids through
porous medium}, Physica, {\bf 1} (1931), 318--333.
\bibitem{Zeldovich}
Zeldovich Ya.,  {\em Theory of Flame Propagation}, National Advisory
Committee for Aeronautics Technical Memorandum 1282 (1951), 39 pp.
\bibitem{KPP}
Kolmogorov A., Petrovskii I., Piskunoff N., {\em Dynamics of Curved
Fronts} (edited by P.Pelce), Academic Press, Boston, 1988, pp.
105-130.
\bibitem{Murray} Murray J. D., {\em Mathematical Biology}, Springer-Verlag, Berlin,
1989.
\bibitem{Kawahara} Kawahara T., Tanaka M., {\em Phys. Letters A,} vol. 97
(1983), 311-314.
\bibitem{Galaktionov2} Galaktionov V., {\em Diff. Int. Equations,} vol. 3
(1990), 863-874.
\bibitem{Clarkson-Mansfields}Clarkson P., Mansfields E., {\em Physica D,} vol. 70
(1993), 250-288.
\bibitem{Cherniha} Cherniha R., {\em J. Math. Anal. Appl.,} vol. 326
(2007), 783-799.
\bibitem{Bar-Yur}Barannyk A., Yurik I., {\em Proc. of the Institute of Mathematics of NAS of
Ukraine,} {vol. 50}, Part I (2004), 29-33.
\bibitem{Nik-Bar}Nikitin A., Barannyk T., {\em Central European Journ. of
Mathematics,} vol. 2 (2005), 840-858.
\bibitem{Ivanova} Ivanova N., {\em Dynamics of PDE,} {vol. 5}, No. 2 (2008), 139-171.
\bibitem{Joseph-Preziosi},  Joseph D.D., Preziozi, L., {\em Review of Modern
Physics,} {vol. 61}, No. 1 (1989), 41-73.
\bibitem{Makarenko}Makarenko A.,  {\em Rep. Math. Physics,} {vol. 46}, No.
1/2 (2000), 183-190.
\bibitem{Makarenko-Moskalkov}Makarenko A.S., Moskalkov M., Levkov S., {\em Phys Lett. A,} {vol. 23} (1997), 391-397
\bibitem{Kar}Kar S., Banik S.K., Ray Sh.,  {\em Jornal of Physics
A: Mathematical and Theoretical, }{vol. 36}, No. 11 (2003),
2771-2780.
\bibitem{VladKu_04}Vladimirov V., Kutafina E., {\em Rep. Math.
Physics,} {vol. 54} (2004), 261--271.
\bibitem{VladKu_05} Vladimirov V., Kutafina E., {\em Rep. Math.
Physics,} {vol. 56} (2005), 421-436.
\bibitem{VladKu_06} Vladimirov V., Kutafina E., {\em Rep. Math.
Physics,} {vol. 58} (2006), 465-476.
\bibitem{VlaMacz_07}Vladimirov V., M\c{a}czka Cz., {\em Rep. Math.
Physics,} {vol. 60} (2007), 317-328.
\bibitem{Kutafina_10}Kutafina E., {\em Journ. of Nonlinear Mathematical Physics,}  vol.
16 (2009), 517-519.
\bibitem{Vladimir_10} Vladimirov V., M\c{a}czka Cz.,  {\em Rep. Math. Physics}, {vol. 65} (2010), 141-156.
\bibitem{Vladimir_09a} Vladimirov V.,  {\em Wave patterns within the generalized
convec\-tion-re\-ac\-tion-dif\-fu\-sion equation}, arXiv:0911.2759v1
[nlin.PS]
\bibitem{GH}Guckenheimer J., Holmes Ph., {\em Nonlinear Oscillations, Dynamical Systems and Bifurcations of Vector
Fields}, Springer, NY, 1987.
\bibitem{Has}Hassard , Kazarinoff, Wan {\em Theory and Applications of the Hopf
Bifurcation}, Springer, NY, 1981.
\bibitem{Vladimir_09}Vladimirov,  {\em Compacton-like solutions of some nonlocal
hydrodynamic-type models,} Proceedings of the IV Workshop "Group
Analysis of Differential Equations and Symmetry and Integral
Systems, October 25-30, 2008, Protaras, Cyprus", pp. 210-225.
\bibitem{Maurin}Maurin K., {\em Analysis,} PWN Publ., Warsaw, 1974.
\bibitem{Simon}Simon B., {\em Sturm Oscillation and Comparison
Theorems,} arXiv:math/0311049 v1 [math. SP]
\bibitem{Idris} Idris I., Biktashev V.N., {\em An analytical approach to initiation of propagating
fronts}, arxiv:0809.0252v1 [nlin.PS]
\bibitem{VlaMacz_11}Vladimirov V., M\c{a}czka Cz., {\em Chaos, Solitons \&
Fractals,} vol. 44 (2011), 677-684.









\end{thebibliography}
\end{document}